\documentstyle[12pt]{article}
\newcommand{\be}{\begin{equation}}
\newcommand{\ee}{\end{equation}}
\newcommand{\etal}{{\it et al.}}
\newcommand{\hmp}{h^{-1}Mpc}

\def\spose#1{\hbox to 0pt{#1\hss}}
\def\ltapprox{\mathrel{\spose{\lower 3pt\hbox{$\mathchar"218$}}
 \raise 2.0pt\hbox{$\mathchar"13C$}}}
\def\gtapprox{\mathrel{\spose{\lower 3pt\hbox{$\mathchar"218$}}
 \raise 2.0pt\hbox{$\mathchar"13E$}}}
\def\inapprox{\mathrel{\spose{\lower 3pt\hbox{$\mathchar"218$}}
 \raise 2.0pt\hbox{$\mathchar"232$}}}

\begin{document}
\centerline{\LARGE MULTIFRACTALITY AS A LINK}
\centerline{}
\centerline{\LARGE  BETWEEN LUMINOSITY AND SPACE DISTRIBUTION}
\centerline{}
\centerline{\LARGE OF VISIBLE MATTER}
\centerline{}
\centerline{}
\centerline{}

\centerline{\Large Francesco Sylos Labini$^{(1,2)}$ and
Luciano Pietronero$^{(2)}$}

\centerline{}
\centerline{10, May, 1995}
\centerline{}

\centerline{\footnotesize ($^1$) Dipartimento di Fisica, Universit\`a di
Bologna, Italy}

\centerline{\footnotesize ($^2$) Dipartimento di Fisica, Universit\`a di Roma
``La Sapienza''}
\centerline{\footnotesize P.le A. Moro 2, I-00185 Roma, Italy}
\centerline{\footnotesize and INFM, Sezione di Roma 1}

\begin{abstract}
We discuss how luminosity and space distribution
of galaxies
are naturally linked in view of their multifractal
properties. In particular we show that the mass (luminosity)
function corresponding to a multifractal distribution in a given
observed volume,
consists of a power law followed  by an exponential cut-off.
This implies that the amplitude $\phi^*$ of the Schechter function
scales with the sample depth, as confirmed by various observational data.
This effect is analogous to the scaling of the
space density due to the fractal nature of the space distribution.
This
has the important consequence that the luminosity function
can be properly defined only in a volume limited   sample.
Also the so-called  "luminosity segregation"
and the concept of bias correspond to a natural consequence of multifractality.
This implies however that they should be considered
from a different perspective with respect to the usual one.
Such a concept allows us to unify the space and luminosity
 distributions as being shaped by a single cause:
multifractality
which should therefore
claim a central stage in theoretical
investigations.
\end{abstract}
\bigskip
\bigskip

{\it Subject headings:} galaxies: clustering - galaxies: structure
- luminosity function - large scale structure of the universe
\newpage

\section{Introduction}

In a well-known review on the galaxy luminosity 
function (LF) Binggeli {\it et al.}(1988)
state that {\em "as the distribution of galaxies 
is known to be inhomogeneous on
all scales up to a least $100 h^{-1} Mpc$, a rich cluster 
of galaxies is like a Matterhorn in a grand Alpine landscape of 
mountain ridges and valleys of length up to 100 Km"}.
The aim of this paper is to consider this point of view in the light 
of the concept of multifractality of the mass distribution. We show how the
main observational aspects of galaxy luminosity and space distributions
are strongly related and can be naturally linked and explained 
as a multifractal (MF) distribution. The concept of
MF is appropriate to discuss physical systems with local 
properties of self-similarity, in which the scaling properties are 
defined by a continuous distribution of exponents.
Roughly speaking one can visualize this property
as having different scaling
properties
for different regions of the system. The fundamental point 
of this paper is that 
not only the pure space distribution of the luminous matter 
is self-similar (fractal), at least up to a certain scale 
(Coleman and Pietronero 1992 - CP92, 
Guzzo \etal 1992; 
Baryshev {\it et al.}, 1994 - BSLMP94;
Pietronero \& Sylos Labini, 1995; 
Sylos Labini {\it et al.}, 1996a - SLGMP96 
Sylos Labini  {\it et al.}, 1996b  
Di Nella {\it et al.}, 1996; 
Sylos Labini \& Amendola, 1996),
 but the whole matter distribution, i.e. 
weighing each point by its mass, is also self-similar. 
This situation 
requires the generalization of the simple fractal scaling 
to a MF distribution in which a  
continuous set of exponents is necessary to
 describe the spatial scaling 
of peaks of different weight
 (mass or luminosity). In this respect the mass and space 
distributions become naturally
entangled with each other.

In {\it section 2} we will briefly 
discuss some characteristic features of the galaxy
distribution such as the space correlation, 
the morphological segregation,
the morphology-density relation, the 
so-called "luminosity segregation" and the 
dwarf and giant galaxies distributions. 
 
The distribution of luminous matter has MF properties: this
result  has been  
derived from the analysis of  various redshift catalogs (CP92, SLGMP96)
 containing information both on space and 
luminous distribution, i.e. position in space and intrinsic luminosity.  
Multifractality naturally links together space and 
luminous (mass) distributions and
provides a mathematical framework to go beyond the standard 
approximation that galaxian luminosities are 
not correlated with spatial location ({\it section 3}). This fact has 
fundamental consequences on the method adopted to determine
 the galaxy LF. In particular we show that the LF has to
be studied only in volume limited samples in order to avoid the 
scaling in LF  shape and amplitude that arises with varying
sample-size. However 
given the limited redshift depth of the available three dimensional 
samples, the shape  of the LF is not 
greatly distorted and it can 
fairly accurately rendered also by magnitude limited samples, 
which are generally used in literature.

After having introduced the main properties of fractal and multifractal 
distributions ({\it section 4}), 
in {\it section 5} we show that  from them
one can derive the shape of the LF
as well as the exponent of the 
two-point correlation function.
Moreover we will relate various observational issues with the 
multifractal behavior of the matter distribution.

Finally in {\it  section 6} we discuss the 
theoretical implications of multifractality 
and in particular its consequences on  the theory of 
galaxy formation.

\section{Galaxy space and luminosity distributions}

We briefly list here the main features of the luminosity
and space distributions of   galaxies 
together with the various 
morphological properties that will be naturally 
embedded in a MF scheme.

1. The  main statistical tool to study
the spatial distribution of galaxies is
 the two point correlation function (CF)
\be
\label {e1}
G(r) = <n(\vec{r_{0}})n(\vec{r}+\vec{r_{0}})> \sim r^{-\gamma}
\ee
where the last equality holds in the case
of fractal distribution, and $D=3-\gamma$ is the fractal dimension.
CP92 by analyzing the CfA1 redshift survey, find that 
the correlation function in Eq.\ref{e1} is a power law 
with exponent $\gamma \sim 1.5$ up to
the sample limit $\sim 20 h^{-1}Mpc$. According to CP92, the CF in 
Eq.\ref{e1}
is the {\it appropriate statistical tool}  that should be 
used in the description of systems with long-range
correlations, rather than the usual $\xi(r)$ (Peebles 1980; Davis \& 
Peebles 1983; see Sec.4 for a detailed discussion). 
Recently there have been available various determinations of the
correlation properties 
in redshift surveys which seem to point to a 
somewhat higher value for 
the fractal dimension. For example
Guzzo \etal (1992) found that in the Perseus-Pisces survey 
(Giovanelli \& Haynes, 1993) 
$D \approx 2 $ up to $30 \hmp$, while they found an 
evidence towards homogenization at larger distances.
This result is confirmed by the analysis of  Sylos Labini \etal 
(1996a, 1996b),
 even though they found 
no evidence towards homogenization and on the contrary they
 extend 
the power law behavior for the density up to $\sim 130 \hmp$.
The discrepancy between 
these results is probably due to the fact that Guzzo \etal (1992)
have used weighing schemes for the treatment of boundary conditions in the 
correlation analysis, that produce an artificial homogenization (CP92).
Moreover Di Nella \etal (1996) by analyzing the LEDA database 
found a similar value for the fractal dimension up to $\sim 150 \hmp$ 
(Fig.1). Another evidence for a higher value of the fractal dimension
is given by Park \etal (1994), who found $\gamma \approx 1$ analyzing 
the CfA2 redshift survey (see also Sylos Labini \& Amendola 1996). 
This value of the correlation exponent
is lower than that found in the CGCC angular catalog, that is 
$ \gamma \approx 1.7$, (Zwicky \etal, 1968).
 The exponent $ \gamma \approx 1.7$ has been obtained also
in the APM galaxy catalog by Collins \etal (1992). 
This difference is 
probably due to the subtle effects that occur in the determination of
the angular projection of a fractal structure (CP92, SLGMP96).
Finally we want to stress that in the case of
 fractal distributions the $\xi(r)$ CF (or the angular CF
 $\omega(\theta)$) has a power law behaviour at  small scales and
 then there is a break that in log-log scale corresponds 
to a steep cut-off. The curved shape of $\xi(r)$, 
or $\omega(\theta)$, makes 
a precise determination of the correlation exponent
very difficult. In particular all the determinations of 
this exponent  by the $\xi(r)$ or $\omega(\theta)$  
are  affected by  a systematic trend towards a larger value
of $\gamma$ (see Sec.4 for a detailed discussion of such an effect).
\smallskip

2. One of the main characteristics of galaxy 
surveys is that one finds that 
groups of galaxies comprise at least $70\%$
of all galaxies not being part of clusters,
and {\em truly isolated galaxies} are very rare.
Tully (1988) by analyzing the {\em Nearby Atlas of Galaxies} 
(Tully and Fisher 1987)
finds that essentially all galaxies can be grouped into clouds and that 
roughly $70\%$ of these can be assigned to groups.
\smallskip

3. Various studies (Eder {\it et al.} 1989, Binggeli {\it et al.} 1990,
Ferguson \& Binggeli, 1994)
of the spatial distribution of {\em dwarf galaxies}
 show that these 
galaxies fall into the structures defined by
the luminous ones and that there is no evidence of segregation
of bright and faint galaxies on large scale:
dwarf galaxies are not more uniformly distributed than giants and 
the dwarfs, as the giants, belong to clouds, groups or clusters and
there is evidence that  the dwarfs fall well into the large scale patterns
suggested by the giants consisting of filaments, walls and arcs.
In particular, there is no evidence for them to fill voids
(Thuan 1987, Bothun {\it et al.}1988).

Disney \& Phillips (1987) pointed out that galaxies of very
{\em low surface brightness} (LSB)
are entirely missed 
for an observational selection effect,
and that what one can see is the {\em "tip of
the iceberg"}.
Bothun et al (1988) concluded that
the patterns of Large Scale Structures appears to be
mostly independent from galaxy surface brightness.
Binggeli {\it et al.}(1988) 
stressed that  most of the galaxies
that are entirely missed because of their low surface brightness seem 
to be also of low total brightness, so that this observational
bias has the effect of a lower cut-off in brightness. 
\smallskip

4. An observation that is particularly 
important from a theoretical point of view 
is  the behavior of the {\em giant-to-dwarf ratio} 
as a function of the local density: we ask whether 
 dwarf galaxies exist in low density regions, 
where
giants are rare, or if they are only found
as satellites of giants so that the 
giants-to-dwarfs ratio does not depend 
on environmental density. There is a clear 
experimental indication that the dwarf-to-giant ratio 
depends on 
the local density (Binggeli {\it et al.}1988, Binggeli {\it et al.}1990,
Ferguson \& Binggeli, 1994). 
Iovino \etal (1993) found that  bright galaxies
are relatively scarce in low density regions, 
while faint spirals are 
poorly present in high-density regions. 
\smallskip

5. Einasto and Einasto (1985) 
found that 
the {\em brightest galaxies} in groups and clusters are brighter 
than in the field by up to 1 magnitude: the brightest galaxies lie
preferentially in dense environments. 
In particular Dressler (1984)
pointed out that the  most luminous 
elliptical galaxies 
usually reside in the clusters 
cores, at local density maxima,  
and are not present in low density fields, 
so that these objects 
seem to be the product of dense environments.
\smallskip

6. The fact that  giant 
galaxies are "more clustered" than the dwarfs 
has been interpreted as corresponding to a larger
value of the amplitude 
of the correlation function for the giants than for dwarfs:
this is the so-called {\it "luminosity segregation"} phenomenon
 (Davis {\it et al.}, 
1988; Iovino \etal, 1993; 
Park {\it et al.}, 1994; Benoist \etal 1996).

On the contrary we show here 
that the segregation of giant galaxies in clusters
arises as a consequence of self-similarity of matter 
distribution, and that in this case the only relevant parameter is the
{\it exponent}
 of the correlation function, while the amplitude is a
spurious quantity that has no direct physical meaning and depends
explicitly on the  sample size.  For a detailed discussion
of the luminosity segregation phenomenon
we refer the reader to Sylos Labini \etal (1996b). 
In Fig.2a we report 
the behaviour of the so-called "correlation length"
$r_0$ (defined as $\xi(r_0)=1$) 
computed in volume limited samples of 
the Perseus-Pisces survey with increasing 
depth $R_s$: the linear scaling is in agreement with 
the fractal behaviour (CP92). Moreover in Fig.2b 
we show $r_0$ computed in volume limited 
samples with the same cut in absolute 
magnitude, but with different depth. 
If the luminosity segregation paradigm were 
to hold one should find
that $r_0$ is independent on sample depth for 
galaxies with the same absolute luminosity, while that is 
clearly  not the case.
On the contrary the 
fractal nature of the galaxy distribution naturally explains
the scaling of $r_0$ with depth. 
\smallskip

7.  There is evidence that galaxies in 
different environments 
 are morphologically  different and may
followed   different evolutionary paths. 
There is  in particular a 
predominance of early type of galaxies in rich clusters:
high density regions are dominated by E 
and S0 galaxies which themselves are hard to find in the field.
Numerous studies have analyzed the variation in the population
fraction and its possible relationship with cluster morphology
(Oemler 1974); the
 {\em morphological segregation} has been studied  
systematically by 
Dressler (1980), who examined the variations in the relative 
fractions of E, S0, and spiral
 galaxies as a function of the {\em local 
density}
and hence quantified the so-called
 {\em morphology-density relation}.
He discovered that the local density 
of galaxies governs the mixture
of Hubble types in any local environment of a cluster,
independently of cluster global parameters like richness or size. 
The correlation of the morphological mix with 
local density is continuous and monotonic.
This behavior has been shown to extend continuously over 6 orders of
 magnitude in space density from rich clusters to low density groups
(de Souza {\it et al.} 1982, Postman \& Geller, 1984). 
 The main features of the   {\em morphological segregation} 
is the decrease in spiral population with increasing
local density and the increase with density of the fraction of S0 
and elliptical galaxies.  

Several authors (Oelmer, 1974; Melnik and Sergent 1977; Dressler 1980; 
Haynes \& Giovanelli  1988; Iovino \etal 1993)
found that the relative abundance
of elliptical, lenticular and spiral galaxies in clusters and their 
peripheries is a function of the local density: 
$80\%$ of   field galaxies are spirals and   $15\%$ of   galaxies
  in rich clusters show spiral structure. 
The morphology-density relation in rich clusters 
is continuous over six orders of magnitude in space density and, 
correspondingly, the galaxian density
is a continuous parameter: the consequence is that the separation
between  the luminosity function and the space density 
is seriously questionable . 
The morphology-density relation is found to hold also
for dwarf galaxies (Ferguson \& Binggeli, 1994).
Recently  Iovino \etal (1993) found clear evidence 
that the morphology and, in a weaker way, luminosity, are two independent 
 parameters  that affect galaxy distribution as a 
function of the local density.
\smallskip

8. The characteristics of morphology segregation can also be
described by a comparison of the 
{\em angular correlation function} for representative 
samples of different morphology.
Davis {\it et al.} (1976) found that elliptical-elliptical 
angular correlation function can be described by a power 
law with a slope significantly steeper than that of  the 
corresponding spiral sample. Moreover the slope of the 
angular correlation function that characterizes the S0-S0 clustering is 
intermediate to other classes. 
 Giovanelli \etal (1986),
 by analyzing the {\em Perseus-Pisces} redshift survey 
found that the slopes of $w(\theta) \sim A \theta^{\beta}$ 
are significantly steeper for early type of galaxies:
for early galaxies $\beta = -0.90$, while for early spirals $\beta = -0.65$ 
and for late spirals $\beta = -0.37$ 

Davis and Djorgovski (1985)  stressed that this result implies 
that the luminous galaxy distribution may not be  a fair tracer of the mass
distribution on any scale. In fact they argued that if the distribution
of light in the Universe is a good tracer of the mass distribution then the
spatial correlation function $\xi(r)$ should be the same for giant and dwarf
 galaxies.
We show in the following that if the galaxy distribution  is multifractal
this apparent paradox is resolved  and one expects that galaxies of different 
masses correlate 
with different exponents of the correlation function.
\smallskip

In the following we will  see that  morphological
segregation is     naturally explained within the 
context of a 
multifractal description, providing in the process 
a quantitative mathematical description of the
phenomena.

\section{Standard analysis of the Luminosity Function}

The differential luminosity function, $\phi(L)$, gives the
probability of finding a galaxy with luminosity 
in the range $[L,L+dL]$ in the unit volume ($Mpc^{-3}$).
In    literature (see Binggeli {\it et al.} 1988 for a review)
one finds several methods to determine the LF for field galaxies and cluster
galaxies. Special emphasis is devoted to the systematic differences in 
 the LF for the various Hubble types. Here we are interested in the 
determination of the {\em general} LF defined as the sum over all Hubble types 
for field galaxies. 
 
Let $\nu(L,\vec{r})$ denote the number of galaxies 
lying in volume $dV$ at $\vec{r}$ that have intrinsic luminosity 
between $L$ and $L+dL$. The main assumption generally used 
(Binggeli \etal, 1988) is that galaxian 
luminosities 
are not correlated with spatial location. 
Under such an hypothesis one can write 
\be
\label{e2}
\nu(L,\vec{r}) dL dV = \phi(L) D(\vec{r}) dL dV
\ee
where $\phi(L)$ gives the fraction of galaxies
per unit luminosity having intrinsic luminosity 
in the interval ($L$,$L+dL$), and $D(\vec{r})$
gives the number of galaxies of all luminosities 
per unit volume 
at $\vec{r}$. 
 
The so-called {\em classical method}
to determine the LF, in addition to the assumptions {\em (1), (2)}
and {\em (3)}, is based
on the hypothesis that the galaxy distribution in the 
samples under analysis has reached 
homogeneity so that the  average density $n_0$
of galaxies in space is constant and well defined.
This method  is highly sensitive to  the spatial inhomogeneities in
the distributions of galaxies
that should distort the shape of the LF. 
For this reason many authors in the past (Felten 1977)
excluded a region of solid angle containing strong "inhomogeneities"
in galaxy distribution as the Virgo cluster.

Given the highly irregular character of  galaxy distribution 
in all the recent redshift surveys 
(Haynes \& Giovanelli, 1988;
Paturel {\it et al.}, 1988;
Da Costa 1994;
Vettolani {\it et al.} 1994), the assumption of constant density
 and homogeneous distribution 
is questionable and, 
 in fact, the amplitude of the LF, that is the average galaxy 
number density, is a strongly fluctuating and not {\em well defined}
quantity in the available samples. 
For this reason all new methods to determine the LF aim at a 
separation between the shape and the amplitude. In particular the so-called
{\em inhomogeneity-independent}
 methods have been developed with the intent to 
determine only the shape of the LF. 
The basic idea is to consider the ratio of galaxies having intrinsic
luminosity between $L$ and $L+dL$ to the total number of
galaxies brighter than $L$. If Eq.\ref{e2} holds then
\be
\label{e3}
\frac{\nu(L,\vec{r})dLdV}{\int_{L}^{\infty}\nu(L',\vec{r})dL'dV} =
\smallskip
\frac{\phi(L)D(\vec{r})dLdV}{\int_{L}^{\infty}\phi(L')D(\vec{r})dL'dV} =
\smallskip
\frac{\phi(L)dL}{\Phi(L)} \sim d\log\Phi(L) \; .
\ee
By differentiating the integrated LF $\Phi(L)$ one 
obtains the differential LF $\phi(L)$.  
This technique allows recovery of 
the shape for the LF undisturbed 
by  space inhomogeneities. 

Usually the LF is assumed to be described by an analytical
function. The most popular is the one proposed by Schechter (1975):
\be
\phi(L)d(L/L^{*}) = \phi^{*}(L/L^{*})^{\alpha} exp(-L/L^{*}) d(L/L^{*})
\ee
where $L^{*}$ is the cut-off, $\phi^{*}$ is the normalization constant 
(amplitude)
and 
$\alpha$ is the exponent.
The LF has been measured by several authors in different 
redshift surveys (De Lapparent \etal, 1986; 
 Loveday \etal, 1992; Da Costa \etal, 1994; 
Marzke \etal, 1994; Vettolani \etal, 1994) and the agreement between
the various determinations in very different volumes is excellent.
The best fit parameters 
are $\alpha =-1.13$ and $M^*_{bj} =-18.70$ (Vettolani \etal, 1994).
The amplitude  $\phi^*$
is the most uncertain parameter of the LF
because of   spatial inhomogeneity 
 in the available samples (De Lapparent \etal, 1986; Da Costa \etal 1994).
We discuss this point in the following.

We will show that  not only the homogeneity assumption is 
inappropriate for the determination of the LF, 
but also that the assumption
in Eq.(2) 
is not satisfied by the actual distribution of
visible matter. 
As the available samples show structures 
as large as the survey depth we will  see 
that  {\em not only the
amplitude $\phi^*$ but also the
cut-off $L^*$ of the LF are dependent on the sample depth}. 
Our essential points will be the following.
The galaxian luminosities  are strongly correlated with their 
positions in space. This  clear observational fact 
can be studied quantitatively with the MF formalism. In particular 
in such a scheme one can 
determine analytically the 
shape  and the amplitude of the
LF, and unify the various observational issues in 
quantitative mathematical scheme.

\section{Essential properties of fractal structures}

In this section we mention the essential properties 
of fractal structures  
and in the following section we introduce the MF formalism.
However in no way these properties are assumed
or used in the analysis itself.
A fractal consists of a system in which more 
and more structures appear at smaller and 
smaller scales and the structures at small 
scales are similar to the one at large scales.
Starting from a
point occupied by an object, we count how 
many objects are present within a volume 
characterized by a certain length scale in
 order to establish a generalized "mass-length" 
relation from which one can define the fractal 
dimension.
We can then define a relation
between $\:N$ ("mass") and $\:r$ ("length") of type (Mandelbrot, 1982)
\be
\label{l2}
N(r) = B\cdot r^{D}
\ee
where the fractal dimension is $D$. 
Fractal structures in physics usually develop 
for length scales limited by a lower and/or an upper cut-off.
In particular there exists  a lower 
scale $r_0$ up to
which the self-similarity holds, 
and  the prefactor $\:B$ 
is  related to the number of elementary objects $N_0$
(galaxies) that are present within $r_0$ (CP92), i.e.
\be
\label{ct}
B= \frac{N_0}{r_0^D}
\ee  
Fractal structures are systems intrinsically irregular at 
all scales, and the self-similarity that characterizes
their properties, implies the absence of regularity 
or analyticity everywhere in the system.
From a mathematical point of view the 
property of self-similarity is associated to
power-law functions
for which the relevant property is
the exponent (fractal dimension); the amplitude provides a
is connection with  the lower cut-offs of the distribution
(CP92, BSLMP94, SLGMP96). 
 
From Eq.\ref{l2} we can readily compute the 
average density $\:<n>$ within a spherical volume
 of radius $R_{s}$ 
\be
\label{l5}
<n> =\frac{ N(R_{s})}{V(R_{s})} = \frac{3}{4\pi } B R_{s}^{-(3-D)} \; .
\ee
From Eq.\ref{l5} we see that the average density 
is not a meaningful concept in a fractal 
because it depends explicitly on the sample 
size $\:R_{s}$. We can also see that for 
$\:R_{s} \rightarrow \infty$ the average density 
$\:<n> \rightarrow  0 $;
therefore a fractal structure is asymptotically  
dominated by voids. 
We can define the {\it conditional average density}, 
that is the density in a spherical shell of area $S(r)$
from an occupied point
\be
\label{l6}
\Gamma (r)= S(r)^{-1}\frac{ dN(r)}{dr} = \frac{D}{4\pi } B r ^{-(3-D)} \; . 
\ee
Usually the exponent  $\:(3-D)$ that defines the decay
of the conditional density is called 
the codimension and it 
is related to the two point correlation function
exponent $\gamma$ as shown in Eq.(1) (CP92).
The conditional average density, 
as given by Eq.\ref{l6}, is well defined in terms of 
its exponent, the fractal dimension. 
Moreover  it is easy to show that the standard two-point correlation function 
in a spherical sample containing a fractal is (CP92)
\be
\xi(r) = [(3-\gamma)/3](r/R_s)^{-\gamma}-1
\ee
so that the so-called "correlation length" is simply a 
linear fraction of the sample size (Fig.2a) 
\be
r_0 = [(3-\gamma)/6] ^{1/\gamma} R_s
\ee
Finally we would like to stress that $\xi(r)$ is a power law 
only for
\be
\frac{3-\gamma}{3} \left(\frac{r}{R_s} \right)^{-\gamma} \gg 1
\ee
hence for $r \ll r_0$: for larger distances there is a clear deviation 
from the power law behaviour due to the definition of $\xi(r)$. 
This deviation, however, is just due to the size of
 the observational sample and does not correspond to any real change 
of the correlation properties. It is clear that if one estimates the
 exponent of $\xi(r)$ at distances $r \ltapprox r_0$, one
 systematically obtains a higher value of the correlation exponent
 due to the break of $\xi(r)$ in the log-log plot. The analysis
 performed by $\xi(r)$ is therefore mathematically inconsistent, if
 a clear cut-off towards homogeneity has not been reached, because
 it gives an information that is not related to the real physical
 features of the distribution in the sample, but to the size of the 
sample itself. 

\section{Multifractal measure}

We now briefly introduce the concept of multifractal measure.
The multifractal picture is a refinement and
generalization of the fractal properties
 (Paladin \& Vulpiani, 1987; Benzi {\it et al.}, 1984, CP92, BSLMP94, 
SLGMP96, Sylos Labini \etal 1996b)
that arises naturally in the case of self-similar distributions.
If one does not consider the mass one has a simple set 
given by the galaxy positions (that we call the {\it support} 
of the measure distribution).
Multifractality instead becomes interesting and a physically 
relevant property when one includes the galaxy masses
and consider the entire matter distribution (Pietronero, 1987; 
CP92). In this case the measure distribution 
is defined by assigning  to each galaxy a weight which is 
proportional to its mass.
The question 
of the self-similarity versus homogeneity of this set can be 
exhaustively discussed in terms of the single correlation exponent
that corresponds to the fractal dimension 
of the support of the measure distribution.
Several authors (Martinez \& Jones, 1990) instead considered
the eventual multifractality 
of the support itself. However the physical implication of 
such an analysis is not clear,
 and it does not add
much to the  question above.

In the more
complex case of MF distributions the scaling properties
can be different for different regions of the
system and one has to introduce a continuous
 set of exponents to characterize the
 system  (the multifractal spectrum).
The discussion  
  presented in the previous section 
was meant to distinguish between homogeneity and scale invariant
properties; 
it is appropriate also in the case of a multifractal.
In the latter case the correlation functions we have considered
would correspond to a single exponent
 of a multifractal spectrum of exponents,
but the issue of homogeneity versus scale invariance
(fractal or multifractal) remains exactly the same.

Suppose that the total volume of the sample
consists of a {\em 3-}dimensional cube of size $L$.
The  density distribution of visible matter is described by  
\be
\label{mf1}
\rho(\vec{r}) = \sum_{i=1}^{N} m_{i} \delta(\vec{r}- \vec{r_{i}})
\ee
where $m_{i}$ is the mass of the $i$-th galaxy
and $N$ is the number of points in the sample
and $\delta(\vec{r})$  is the Dirac delta function.
We assume that this distribution 
corresponds to a measure defined on the set of
points which have the correlation properties described by Eq.\ref{l6}.
It is possible to define the dimensionless 
normalized density function
\be
\label{mf2}
\mu(\vec{r}) = \sum^{N}_{i=1} \mu_{i} \delta(\vec{r}-\vec{r}_{i})
\ee
with $\mu_{i} = m_{i}/M_{T}$ and $M_{T} = \sum^{N}_{i=1} m_{i}$,
the total mass in the sample.
We divide this volume into boxes of linear size $l$. We label each
 box by the index $i$ and construct for each box the function
\be
\label{mf3}
\mu_{i}(\epsilon)  = \int_{{\it i-th box}} \mu(r)dr
\ee
where $\epsilon = l/L$ and $0 < \mu_i<1$.
The definition of the box-counting fractal dimension is
\be
\label{mf4}
\lim_{\epsilon \rightarrow 0} \mu_{i}(\epsilon) \sim \epsilon^{\alpha (\vec{x})}
\ee
where $\alpha (\vec{x})$ is constant and equal to $D$ 
in all the occupied boxes in the case of a simple fractal.
%
This exponent 
fluctuates widely
with the position $\vec{x}$ in the 
case of MF.
 In general we will find
 several boxes with a measure that scales with the
same exponent $\alpha$. These boxes form a fractal subset
 with dimension $f$ that  depends on the 
exponent $\alpha$, i.e. $f=f(\alpha)$.
Hence the number of boxes that have a measure $\mu$ that scales
with exponent in the range [$\alpha , \alpha + d\alpha$]
varies with $\epsilon$ as
\be
\label{mf5}
N(\alpha, \epsilon)d\alpha \sim \epsilon^{- f(\alpha )} d\alpha.
\ee
 The function $f(\alpha)$ is usually (Paladin \& Vulpiani 1987)
a single humped
function with the maximum at $max_{\alpha} f(\alpha) = D$, 
where $D$ is the dimension of the support. In the case
of a single fractal, the function $f(\alpha)$ is reduced to
a single point: $f(\alpha) = \alpha = D$

In order to analyze  the mass distribution of galaxies,
obviously one needs to know the density distribution
 $\rho(\vec{r})$. The mass of each galaxy may be related to its total
luminosity in a simple way
\be
M= k(i) L^{\beta}
\ee
where $k$ is the mass to light ratio and depends on the galaxy 
morphological type $i$. With respect to the MF properties,  
$k$ plays a little role because the important quantity 
is the range of galaxy mass, which can be as large
as a factor $10^6$ or more. Therefore a modification of $k$ 
produces small effects on a logarithm scale. The exponent $\beta$
is  more important,
 and here we assume (Faber and Gallagher 1979) 
that $\beta \approx 1$.
However a different value of $\beta$ 
 should not change the MF nature of the mass
distribution, if it is present in the sample, but only the
parameters of the spectrum.

From a practical point of view one does not
determine directly the spectrum of exponents $[f(\alpha), \alpha]$;
 it is more convenient to compute   its Legendre
transformation $[\tau(q),q]$ given by
\be
\left\{
\begin{array}{l}
\label{mf12}
\tau(q)=q\cdot \alpha(q)-f(q)\\
\frac{d\tau(q)}{dq}=\alpha(q)
\end{array}
\right.
\ee
In the case of a simple fractal one has
 $\alpha=f(\alpha)=D$.
In terms of the Legendre transformation
this corresponds to 
\be
\tau(q) = D(q-1)
\ee
i.e. the behaviour of $\tau(q)$ versus $q$ is a straight line with coefficient 
given by the fractal dimension.

The analyses 
carried out on CfA1 (CP92) and Perseus-Pisces
 (Sylos Labini \etal, 1996b)
 redshift surveys
provide unambiguous
evidence for a MF behavior as shown by the 
non linear   
behaviour of $\tau(q)$ in Fig.3.

\subsection{Multifractal measure distribution}

We have seen in the previous section the basic formulae
that describe 
the scaling properties of a multifractal measure (MF).
Suppose now we have a MF sample 
in a well defined volume $V$, 
and we want to study the behavior of the number 
of boxes with measure in the range 
$\mu$ to $\mu+d\mu$, having fixed the 
partitioning of the measure 
with boxes of size $\epsilon$.
By  changing variables and using Eq.\ref{mf4},
then the measure distribution (Eq.\ref{mf5}) becomes
\be
\label{f3}
N_{\epsilon}(\mu) d\mu \sim \epsilon^{-f(\alpha(\mu))}
\frac{1}{|\log(\epsilon)|}
\frac{d\mu}{\mu}.
\ee
From this equation we can see that 
the distribution of the 
measure, at fixed resolution $\epsilon$, does not scale as a power law in 
$\mu$, because the exponent $f(\alpha(\mu))$ 
is a complex function of $\mu$.
The self-similarity of the distribution is recovered 
by looking at the measure distribution as a function of the scale 
$\epsilon$. 

Suppose we fix the dimension of the box at the scale 
$\epsilon$: for example, we can suppose that 
this can be the galactic scale, or the cluster  scale.
The function $N_{\epsilon}(\mu)$ is 
bell-shaped and convex 
with a maximum corresponding
 to the point at which
\be
\label{f4}
\frac{\partial N_{\epsilon}(\alpha)}{\partial\mu} = 
\frac{\partial N_{\epsilon}(\alpha)}
{\partial\alpha}\frac{\partial \alpha}{\partial\mu} 
= -(f'(\alpha)+1) N_{\epsilon}(\alpha) \frac{1}{\mu} 
= 0
\ee
this condition corresponds
\be
\label{f5}
\left( \frac{\partial f(\alpha)}{\partial\alpha}\right)_{\alpha_{c}} = -1.
\ee
The maximum of $N_{\epsilon}(\mu)$ fixes the most probable
value of $\mu$.
 Well beyond this
maximum the function can be well fitted 
by a power law. 
In practice this is the only observable part of
the measure distribution in the case of galaxies
because the higher values of $\alpha$ correspond to
the smallest galaxies that are not present in the sample (CP92).
For still higher values
of $\mu$ the function 
shows an exponential-like  decay. The tail is fixed by the point
at which the derivative Eq.\ref{f5} has a maximum.
This happens for
$\alpha = \alpha_{min}$, namely at the value corresponding to the box that contains the maximum measure (i.e. the strongest
singularity)
\be
\label{f6}
\mu^{*}  \sim \epsilon^{\alpha_{min}}.
\ee
In order to compute the exponent
characterizing the leading power law behavior 
we study the derivative of $\log(N_{\epsilon}(\mu))$ with 
respect to $\log(\mu)$.
By performing the logarithmic derivative of Eq.\ref{f3} we obtain
\be
\label{f7}
\frac{\partial \log (N_{\epsilon}(\mu))}{\partial\log(\mu)} =
- \left( \frac{ \partial f(\alpha) } {\partial \alpha} +1\right).
\ee 
We can try to fit Eq.\ref{f3} with a power law function of $\mu$, plus an
exponential tail. From Eq.\ref{f7} 
we can define an effective exponent $\delta$, which depends 
explicitly on $\mu$. This implies that the power 
law approximation can be considered as a local fit 
\be 
\label{f8}
\delta =- \left(\frac{\partial f(\alpha)}{\partial \alpha} +1\right).
\ee
This leads to $\delta = 0$ for $\alpha=\alpha_{c}$ and
$\delta=$-1 for $\alpha_{0}$ such that $f(\alpha_{0})=D(0)$. 
Locally we can expand $f'(\alpha)$ in power series of $\mu$,
so that
the Measure Distribution (MD) 
of Eq.\ref{f3} in a certain range of $\mu$,
is well fitted by a power law function 
with a cut-off
\be
\label{f9}
N(\mu) \sim \mu^{\delta} e^{-\frac{\mu}{\mu^{*}}}.
\ee
The exponent $\delta$
depends on the shape of the derivative of $f(\alpha)$,
as well as on the value of $\alpha$
around  which one develops $f(\alpha)$. We shall see, 
with the help of computer simulations 
that the value of $\delta$
is usually 
in the range $[-2,-1]$ for a wide range of 
$f(\alpha)$ spectra.

\subsection{Numerical simulations}
In order to examine a more general case 
we now 
consider a  
 random multiplicative measure
 generated by a random 
fragmentation process using 
$m$ normalized generators $\chi^{(\gamma)}$ and $\gamma =1,..,m$ 
in the $d$-dimensional Euclidean space (Fig.7).
The analytical shape of the measure distribution can be computed 
from the knowledge of $f(\alpha)$,
determined by the Legendre transformation of $\tau(q)$ (computed from the 
$\chi^{(\gamma)}$).
It is interesting to see how a MF distribution 
naturally leads to the various morphological properties
that we have discussed in Sec. 2.
\bigskip

{\it (1) The two point number-number correlation function}
\smallskip

We have seen that the characteristics of a MF distribution is 
the $f(\alpha)$ spectrum of exponents.
The relation of the properties of this distribution with the usual
correlations analysis 
is straightforward. The exponent that 
describes the power law behaviour of the 
number density corresponds
to the support of the MF distribution and 
therefore it is related to the maximum of $f(\alpha)$
 (see Sec. 5.1).
\bigskip

{\it (2) Luminosity function}
\smallskip

In Fig.(8) we show the behaviour of the tail 
of $P(\mu)$ computed for the random MF distribution
shown in Fig.7.
 It is possible to fit this curve with a 
Schechter-like function (Eq.\ref{f9}): the best fit gives
$\delta=-1$ in agreement with Eq.\ref{e40}. 
We can also see  that 
at small $\mu$ the power law behaviour is 
a little bit steeper than in the intermediate region.

Consider a  portion of a MF distribution in 
a spherical volume of radius $R$. 
The number of singularities of type $\alpha$ 
in the range $[\alpha, \alpha+ d \alpha]$, is given by 
Eq.\ref{mf5}. The corresponding average space density scales therefore as 
$<\nu(\alpha,R)> \sim R^{(f(\alpha)-3)}$. This relation implies 
that the space density is a complex function 
of $\alpha$ and $R$.  Although not strictly valid, 
a separation between a space and a 
  luminosity distribution is
 useful in the analysis of real catalogs.
We approximate such a separation as follows.
Integrating  Eq.\ref{mf5} 
in $d\alpha$ we obtain that 
the  total number of  boxes of size $\epsilon$, divided for the volume,
will be
\be
n(R) =  \frac{N(\epsilon(R))}{V} = \frac{3}{4\pi} B R^{D(0)-3}
\ee
where the last equality follows from Eq.\ref{l5}.

The probability that a certain box (at a given scale $\epsilon$)
has the measure in the range ($\mu$, $\mu+d\mu$)
 is determined by 
Eq.\ref{f3} (or Eq.\ref{f7}). Hence  
the average number of boxes for unit measure 
and unit volume can be written as 
\be
\label{new2}
<\nu(R,\mu)> = \frac{3}{4\pi} B R^{D(0)-3} \mu^{\delta} 
e^{-\frac{\mu}{\mu^*}}
\ee
This equation can be read as the {\it average} probability
of having a galaxy of a certain luminosity and in a certain volume,
in a MF Universe. 

The amplitude of $\nu(R,\mu)$ is related 
to the lower cut-offs 
of the distribution and the size of the sample volume, and
therefore
has no special physical meaning.
We stress in the following (see {\it (5)}) 
that the shape of the LF is not completely independent 
on the size of the sample volume.
This implies that the approximation of Eq.\ref{new2}
 does not strictly hold, because it does not 
consider the correlation between 
space and luminosity distribution, that for MF 
we know to be an important feature 
(see {\it (6)}).
Nevertheless we stress that Eq.\ref{new2} can be used in practice 
in the analysis of
real redshift surveys, with great  accuracy.
In fact, the result 
of Eq.\ref{new2} has been obtained under the approximation of
Eq.\ref{f9}, while in the more general case the cut-off $\mu^*$
depends explicitly on the sample size. 
However this dependence is very weak in the available samples.

In summary, in 
order to study a MF distribution the
three dimensional volume must be well defined:
 only in such a kind of volume
one may define the scaling properties of the MF. This implies that 
an understanding of the distortions of the LF shape requires that 
 volume limited samples should be used,
rather than  magnitude limited ones.
Several authors, by analyzing the mean density in redshift surveys
(i.e. the amplitude of the LF), concluded that 
the samples are not large enough to be {\it fair}
because the fluctuations are too large (Da Costa {\it et al.}, 1994;
Marzke {\it et al.}, 1994). Our conclusion is that in magnitude limited 
samples it is still possible to use the inhomogeneity-independent 
technique to determine, if the galaxy distribution is MF, the 
shape of the LF, but its is certainly not possible to recover the 
amplitude of the LF, that is related to the space distribution 
via the average density. In this respect one should consider 
a volume limited sample and normalize for the global
luminosity selection that is related to such volume limited 
sample (SLGMP96, Sylos Labini \etal 1996b). 
\bigskip

{\it (3) Morphological and 
luminosities properties: a new interpretation of "luminosity-segregation"}
\smallskip

Massive  galaxies are mostly found in rich clusters
while field  galaxies are 
usually spirals or gas rich dwarfs (see Sec.2). 
These observational properties 
are consistent with multifractality, i.e. with
the self-similar behaviour of the whole 
matter distribution. In Fig.4 we show the case of a
deterministic MF while Fig.7 corresponds  to
a random MF. The largest peaks are located in the largest
clusters. 
For the self-similarity
each point of the structure belongs to a cluster or to a group of galaxies, 
because a certain portion of the 
fractal distribution is always made of smaller and smaller structures.
Moreover the observations that the dwarf 
and low-surface-brightness galaxies do not fill the voids, 
is consistent with the fact 
that the galaxy distribution continues 
to be fractal 
even for the lowest peaks of the MF.
In this picture the giant-to-dwarf ratio
depends on the environmental density.
In fact, the dwarf galaxies can belong
 to the rich clusters where the giants lie,
but they can also be in small groups.
The morphological-segregation can be seen 
as the  self-similar character 
of the matter distribution. Multifractality is a 
description that can be useful for the statistical characterization 
of the system, but, of course, it cannot explain in detail various evidences 
which require a more appropriate morphological analysis.

From our result we can conclude that the fractal nature of galaxy
distribution in the available samples, can account
for the scaling of $r_0$ with sample depth. In this sense 
the luminosity segregation, intended as a different clustering 
properties of brighter and fainter galaxies in terms of the 
amplitude of the standard correlation function (Davis \etal, 1988;
Park \etal, 1994; Benoist \etal, 1996) has no experimental support.
The linear dependence of $r_0$ on the sample size can be completely
 explained by the fractal nature of galaxy distribution (see also
CP92, BSLMP94, Sylos Labini \etal, 1996; Di Nella \etal, 1996; 
Sylos Labini \& Amendola 1996). The correct perspective to
describe the different clustering of brighter and fainter galaxies 
is  the MF picture, for which we have given ample evidences, 
and 
that implies that massive 
are mostly into large clusters 
as observed. The quantitative characterization of such a phenomenon
is therefore in terms of the exponent of the correlation function 
rather than its amplitude. In particular the brighter galaxies
should have a greater correlation exponent than the fainter ones
(see {\it (4)}).
\bigskip

{\it (4) Multifractal spectrum and multiscaling}
\smallskip

In   Sec.4 we have introduced the MF spectrum $f(\alpha)$ 
and now we clarify its basic properties. Multifractality implies
that if we select only the largest peaks in the measure distribution,
the set defined by these peaks may have different fractal dimension 
than the set defined by the entire distribution.
One can define a cut-off in the measure and consider only those
singularities that are above it. If the distribution
is MF the fractal dimension decreases as the cut-off increases.
We note that, strictly speaking, the presence of the cut-off
can lead (for a certain well defined value of the 
cut-off itself) to the so-called {\it multiscaling}
behavior of the MF measure (Jansen {\it et al.}, 1991).
In fact, the presence of a lower cut-off 
in the calculation of the generalized correlation function affects
the single-scaling regime of $\chi(\epsilon,q)$
for a well determined value of the cut-off $\alpha_{cut-off}$
such that $\alpha_{cut-off} < \alpha_c$, 
and this function exhibits a slowing varying exponent proportional to the 
logarithm of the scale $\epsilon$. However some authors
(Martinez {\it et al.}, 1995)
misinterpret the multiscaling of a MF distribution
as the variation of the fractal dimension with the 
density of the sample.

The fractal dimension $D$ of the support corresponds to the 
peak of the $f(\alpha)$-spectrum 
and raising the cut-off implies a
drift of $\alpha$ towards $\alpha_{min}$
so that $f(\alpha)<D$ (Fig.5).
This behavior can be connected with the different
correlation exponent found by the angular correlation function
for the elliptical, lenticular and spiral 
galaxies (see Sec. 2). In particular the observational
evidence is that the correlation exponent is higher for elliptical
than for spiral galaxies: this trend is compatible with a 
lower fractal dimension for the more massive galaxies than for 
the smaller ones, in agreement with a  MF behaviour.
\bigskip

{\it (5) Scaling of the maximum mass}
\smallskip

As shown in CP92 an important feature of the $f(\alpha)$ 
spectrum is represented by the value at $\alpha_{min}$: 
$f(\alpha_{min})$. This exponent corresponds to the scaling of the 
maximum singularity
\be
\label{sc1}
\mu_{max}(\varepsilon) \sim \varepsilon^{\alpha_{min}}
\ee
where $\mu_{max}(\varepsilon)$
id the maximum measure among all the boxes corresponding to a 
gridding of size $\varepsilon$.
The corresponding maximum density $\rho_{max}$
is therefore $\rho_{max} \sim \mu_{max}/\varepsilon^{3}$
where $\varepsilon \sim 1/R_S$ and $R_S$ is the total
size of the system.
Under the physical assumption that the maximum 
mass $M_{max}$ is related to the maximum density one can conclude that:
\be
\label{sm2}
M_{max} \sim R_s^{3-\alpha_{min}}
\ee
i.e. that the maximum galaxy mass we can observe in a certain sample is
related to the size of the sample itself. To study 
such an effect the sample size should be varied over a large range 
of length scales. In practice the depth of the VL samples 
that can be extracted from the available redshift surveys does not allow
one to detect the scaling implied by  Eq.\ref{sm2}
so that the inhomogeneity-independent  
method in 
magnitude limited samples 
remains the most suitable to study the LF. 
\bigskip

{\it (6) Correlation between space and luminosity distribution}
\smallskip

In Sylos Labini \etal (1996b)
 we have performed a test to check the MF
nature of the observed Perseus-Pisces catalog.
We have randomized 
the absolute magnitudes of the galaxies,
i.e. we have fixed the galaxy position and we have assigned
to each galaxy an absolute  magnitude
chosen at random among all the other galaxies.
 By doing this one destroys
the correlations between the spatial locations and
magnitudes of galaxies. 
 As shown in
Fig.9, in this case the dependence of
$\:\tau(q)$ versus $\:q$ is
linear, as for a
simple fractal, with slope $\approx 2$.
Instead in Fig.3 we show that the original
distribution as a curved shape for $\tau(q)$.
This result shows that the locations of galaxies 
are intrinsically correlated with their luminosities, i.e.
the existence of a luminosity-position correlation. 
The MF framework provides a mathematical tool to study 
such a distribution.

\section{Discussion and Conclusions}
In this paper we have shown that it is possible to  
frame  the main properties of the galaxy space and luminosity 
distribution  in a unified scheme, by 
using the concept of multifractality (MF).
In fact, the continuous set of exponents
$[\alpha,f(\alpha)]$ that describes a MF distribution
can characterize completely 
the galaxy distribution when one considers 
the mass (or luminosity) of galaxies in the analysis.
In this way many observational evidences are linked together
and arise naturally from the self-similar properties
of the distribution.

Considering a MF  distribution, the usual power-law
space correlation properties correspond just to
 a single exponent of the $f(\alpha)$ spectrum:
such an exponent simply describes the space distribution of the support
of the MF measure. 
Furthermore the shape of the luminosity function (LF),
i.e. the probability of finding a  galaxy of a certain luminosity
per unit volume, is related to the $f(\alpha)$
spectrum of exponents of the MF.
We have shown that, under MF conditions, the LF is well 
approximated by a power law function with an exponential tail.
Such a function corresponds to the Schechter LF observed in real
galaxy catalogs. In this case the 
shape of the LF is almost independent on the sample size.
Indeed we have shown that a weak dependence on sample size  
is still present because the cut-off of the Schechter function 
for a MF distribution turns out to be related to the sample 
depth: $L^*$ increases with sample depth.
In practice as this quantity is a strongly fluctuating one,
in order to study its dependence on the sample size one should have a 
very large sample and should vary the depth over a 
large range of length scales.
Given this situation 
a sample size independent  shape of the LF can be well 
defined using the inhomogeneity-independent method in magnitude limited 
samples.
Indeed such a technique has been introduced
to take into account the highly irregular 
nature 
of the large scale galaxy distribution. For example 
a fractal distribution is non-analytic in each point and it is not
possible to define a meaningful average density.
This is because the intrinsic fluctuations that characterize 
such a distribution can be large as the sample itself, and the 
extent of the largest structures is limited only by the 
boundaries of the available catalogs.

Moreover if the distribution is  MF, the amplitude of the 
LF depends on the sample size as a power law function.
To determine the amplitude of the LF, as well as the 
average density, one should have a well defined volume limited sample, 
extracted from 
a three dimensional survey. We refer the reader to SLGMP96
for a detailed discussion on the determination of the 
average density that is related to the 
determination of the lower cut-off of the fractal distribution and to the 
subtle problems related with the finite size effects.

These results have important consequences from a theoretical point of view.
In fact, when one deals with self-similar structures the relevant 
physical phenomenon that leads to the scale-invariant structures is
characterized by the {\it exponent} and {\it not the amplitude} of the 
physical quantities that characterizes such distributions.
Indeed, the only relevant and meaningful quantity is the 
exponent of the power law correlation function
(or of the space density), while the amplitude of the correlation 
function, or of the space density and of the LF, is just 
related to the sample size and to the lower cut-offs of the distribution.

The geometric self-similarity has deep implications for the 
non-analyticity of these structures. In fact, analyticity or
regularity would imply that at some small scale 
the profile becomes smooth and one can define 
a unique tangent. Clearly this is impossible in a self-similar
structure because at any small scale a new structure appears and the 
distribution is never smooth. Self-similar structures are therefore
intrinsically irregular at all scales and correspondingly one 
has to change the theoretical framework into one which is 
capable of dealing with non-analytical fluctuations. This means going
from differential equations to something like the 
Renormalization Group to study the exponents.
For example the so-called "Biased theory of galaxy formation"
(Kaiser, 1984) is implemented considering the evolution of 
density fluctuations within an analytic Gaussian framework, 
while the non-analyticity of fractal fluctuations 
implies a breakdown of the central limit theorem which is the 
cornerstone of Gaussian processes (Pietronero \& Tosatti 1986, 
CP92, BSLMP94).

In this scheme {\it the space correlations and the luminosity
function are then two aspects of the same
phenomenon, the MF distribution of visible matter}.
The more complete and direct way to study such a 
distribution, and hence at the same time the space 
and the luminosity properties, is represented by the 
computation of the MF spectrum of exponents.
This is the natural objective 
of theoretical investigation in order 
to explain the formation and the distribution of galactic
structures. In fact, from a theoretical point of view one would like
 to identify the dynamical processes that can lead to such a 
MF distribution. As a preliminary step in this direction we have 
developed a simple stochastic model (Sylos Labini \& Pietronero, 1995a;
Sylos Labini \& Pietronero, 1995b)
in order to study which are the fundamental physical effects that 
lead to such a MF structure in an aggregation process. This is a very complex
problem but to study it correctly one has to use the appropriate concepts
and statistical tools.

If a crossover towards homogeneity would eventually be detected, 
this would not change the above discussion but simply introduce 
a crossover into it. The 
(multi)fractal nature of the observed structures would in any case, 
require a change of theoretical perspective.

\section*{Acknowledgements}
It is a pleasure to thank M. Haynes and R. Giovanelli for having 
given us
the possibility of analyzing the Perseus-Pisces redshift survey.
We thank L. Amendola, M. Costantini, H. Di Nella, A. Gabrielli, 
M. Montuori, G. Paladin, A. Vulpiani, P. Vettolani and G. Zamorani
for useful discussions and constructive comments.

\section*{Appendix: The case of the Binomial  multifractal}
We discuss here a simple example
of self-similar
multifractal distribution for which all the properties
can be derived analytically. This will provide a useful playground
for the discussion of the real matter distribution.
The binomial measure is defined by the following process:
we start with constant density $\rho =1$ defined in 
 the unit interval $[0,1]$ (the so-called support of the distribution).
The first iteration consists of dividing the unit interval 
into two equal pieces, one having weight $\mu_{1}$  
and the other one 
with weight $\mu_{2}$ such that $\mu_{1}+\mu_{2}=1$.
During the next iteration we apply the same procedure 
for the two subintervals:
 after an infinite number of iterations 
we are left with a well defined binomial measure
(Fig.4). 
After $k$ iterations
the probability of having a singularity with measure 
(Pietronero \& Siebesma, 1986)
\be
\label{e30}
\mu=\mu_{1}^{i} \mu_{2}^{k-i} 
\ee
is
\be
\label{e31}
P(k,i) = \frac{1}{2^{k}}\left(\begin{array}{c} k\\i \end{array}\right) 
\ee
with $i=1,...,k$.
Considering the continuous limit of the binomial for $k\rightarrow\infty$
we have:
\be
\label{e32}
P(k,i)di \sim \frac{1}{2\pi\sigma^{2}} e^{-\frac{(i-<i>)^{2}}{2\sigma^{2}}} di
\ee
with:
\be
\label{e33}
<i> = k/2 
\ee
\be
\label{e34}
\sigma^{2} = k/4
\ee
Inverting Eq.\ref{e30}, we obtain 
\be
\label{e35}
i = a  \log(\mu) + b
\ee
where 
\be
\label{e36}
a= \frac{1}{\log(\mu_{1})-\log(\mu_{2})}
\ee
\be
\label{e37}
b=- k a \log(\mu_{1})
\ee
By changing the variables in Eq.\ref{e31}, using Eq.\ref{e35},
 we obtain
\be
\label{e38}
P(\mu)d\mu \sim \frac{1}{2\pi\sigma^{2}} 
e^{\frac{-(a\log(\mu)+c)^{2}}{2\sigma^{2}}} \frac{a}{\mu} d\mu
\ee
\be
\label{e39}
c= b - <n>
\ee
From Eq.\ref{e38} we can see that 
\be
\label{e40}
P(\mu) \sim \mu^{-1}
\ee
if the following inequality is satisfied
\be
\label{e41}
\frac{a^{2}(\log(\mu))^{2}}{2\sigma^{2}} < 
\left(1+\frac{|ac|}{\sigma^{2}}\right) |\log(\mu)| \theta
\ee
or 
\be
\label{e42}
|\log(\mu)| < \frac{2\sigma^{2}}{a^{2}} \left(1 +
 \frac{|ac|}{\sigma^{2}}\right) \theta
\ee
where we have introduced the 
parameter $\theta$ to define
quantitatively  the precision of 
the approximation.
For example for $\theta = 0.1$ then Eq.\ref{e40} holds 
to about 
$10\%$ accuracy 
(Fig.5).  Beyond this region the function has a decay that
can be fitted by an exponential tail.

\newpage

\section*{References}
\begin{itemize} 

\item Baryshev Y., Sylos Labini F., Montuori M. \& Pietronero L.,
1994 Vistas in Astron. 38, 419 (BSLMP94)

\item Benoist C., Maurogordato S., da Costa L.N., Cappi A. and 
Schaeffer R., 1996 preprint

\item Benzi, R., Paladin, G.,Parisi, G. and Vulpiani, 
A. 1984, J.Phys. A 17,3521

\item Binggeli B., Sandage A., Tamman G.A., 1988 Ann. Rev. A\&A  26, 509

\item Binggeli B., Tarenghi M., Tamman G.A. 1990 A\&A 228,42

\item Bothun G. D., Beers T.C., Mould J.R. and Huchra J.P., 1988 Ap.J 308, 510

\item Coleman, P.H. \& Pietronero, L.,1992 Phys.Rep. 231,311 (CP92)

\item Collins C. A., Nichol R.C and Lumsden S.L.  1992 MNRAS 254, 295

\item Da Costa  L., Geller M.J., Pellegrini P.S., 
Latham D.W., Fairall A.P., Marzke R.O., Willmer C.N.A., Huchra J.P.,
Calderon J.H., Ramella M. and Kurtz M.J.  1994 Ap.J. 208, 13

\item Da Souza  R.E., Capela H.V., Arakaki L., Logallo C., 
1982 Ap.J. 263, 557

\item Dalton G. B., Croft R.A., Efstathiou G., Sutherland W.J.,
Maddox S.J. and Davis M.  1994 Ap.J. 271, L47 

\item Davis M. \& Geller M., 1976 Ap.J. 208, 13

\item Davis M. \& Peebles, P. J. E.  ., 1983 Ap.J. 267, 465

\item Davis M. \& Djorgovski S., 1985 Ap.J. 299, 15

\item Davis M. , Meiksin A., Strauss M.A., da Costa L.N. and 
Yahil A., 1988 Ap.J. 333, L9 

\item De Lapparent V., Geller M. \& Huchra J., 1989 Ap.J. 343, 1

\item Di Nella H., Montuori M., Paturel G., Pietronero L.
 \& Sylos Labini F., 1996 Astron.Astrophys.Lett., 308, L33

\item Disney M.  \& Phillips, S. 1987 Nature 329, 203 

\item Dressler A., 1984 Ann.Rev. A\& A 313, 42

\item Dressler A., 1980 Ap.J. 236, 351

\item Eder J. A. , Schomebert J.M., Dekel A. and Oelmer A. Jr.
 1989 Ap.J. 340, 29

\item Einasto M. \& Einasto J., 1987 MNRAS 226, 543

\item Faber S,M. \& Gallagher, J.S. 1979 Ann Rev A\&A 17, 135

\item Felten J., 1977 AJ 82, 861

\item Ferguson H.C. \& Binggeli B., 1994 Ann. Rev. A\& A 

\item Giovanelli R. , Haynes M.P. and Chincarini G., 
1986 Ap.J. 300, 77

\item Giovanelli R. \& Haynes M. 1993 AJ 

\item Guzzo  L. , Iovino A., Chincarini G,   Giovanelli R. , Haynes M.P. 
1992 Ap.J. 382, L5

\item Haynes M. P. \& Giovanelli R. 1988 in "Large scale motion in 
the 
Universe", ed. Rubin V.C and Coyne G., Princeton University Press,
Princeton

\item Iovino  A.,  Giovanelli R. , Haynes M.P., Chincarini G
 Guzzo  L. 1993 MNRAS 265, 21 

\item Jensen M.H., Paladin G., Vulpiani A. 1991 Phys.Rev.Lett. 67, 208

\item Kaiser N.1984, Ap.J. 284, L9

\item  Loveday J., Peterson B.A., Efstathiou G., Maddox S.J.
1992 Ap.J. 390, 338 

\item Mandelbrot, B., 1982 "The fractal geometry of nature",
 Freeman New York

\item Martinez V.J. \& Jones B. 1990 MNRAS 242, 517

\item Martinez  V. J., Paredes S., Borgani S. and Coles P.
1995 Science 269, 1245

\item Marzke R. O., Huchra J., Geller M. 1994 Ap.J. 428, 43

\item Melnick J., \& Sergent W.L.W. 1977 Ap.J. 215, 401

\item Oelmer A. 1974 Ap.J. 194, 1

\item Paladin, G., Vulpiani,A. 1986 Phys. Rep. 156, 147

\item Paturel G, Bottinelli L., Gougueneheim L., Foque P.
1988 A\&A 189, 1

\item Park C., Vogeley M.S., Geller M.J. and Huchra J.P.
1994 Ap.J. 431, 569

\item Pietronero, L., \& Tosatti, E. (eds) 1986, "Fractals in Physics"
 (North-Holland; Amsterdam)

\item Pietronero, L.\& Siebesma A. P. 1986 Phys.Rev.Lett. 57, 1098

\item Pietronero L. 1987 Physica A, 144, 257

\item Pietronero L. \& Sylos Labini F., 1995 in "Birth of the Universe
and fundamental physics" F. Occhionero ed., Springer Verlag, p. 17

 \item Postman M. \& Geller M., 1984 Ap.J. 281, 95

\item Schechter P., 1976 Ap.J. 203, 297
 
\item Sylos Labini F. \& Pietronero L. 1995a in  "Birth of the Universe
and fundamental physics" F. Occhionero ed., Springer Verlag, p. 317.
 
\item Sylos Labini F. \& Pietronero L. 1995b  
Astrophys. Lett. and Comm., 36, 49

\item Sylos Labini, F., Montuori, M., Pietronero, L., 1996b 
Physica A, in print 

 \item Sylos Labini, F., Gabrielli A., 
Montuori, M., Pietronero, L., 1996
Physica A, 226, 195

\item Sylos Labini F. \& Amendola L., 1996 Astrophys. Lett. and
 Comm., in print 

\item Thuan T. X. 1985 Ap.J. 299, 881

\item Tully R.B. 1988 in "Large scale motion in the
Universe", ed. Rubin V.C and Coyne G., Princeton

\item Tully R.B. \& Fisher J. R., 1987 "Nearby Galaxies Atlas"
Cambridge: Cambridge University Press

\item Vettolani P. {\it et al.} 1994 Proc. of Scloss Rindberg Workshop
"Studying the Universe with Clusters of  Galaxies"

\item Zwicky F., Herzog E., Wild P., Karpowicz M. and Kowal C.T.
1961-1968, Catalogues of Galaxies and Clusters (6 vols) 
(California Institute of Technology)
\end{itemize}
\newpage

\section*{Figure captions}
\begin{itemize} 

\item Fig.1 The average conditional density $\Gamma(r)$ for VL samples of
several redshift surveys (normalized to the luminosity
selection of the VL sample): 
Perseus-Pisces (Sylos Labini \etal 1996b, 
SLGMP96), LEDA (Di Nella \etal 1996),
CfA1 (CP92). The reference line has a slope 
$-\gamma=-1$ ($D \approx 2$).

\item Fig.2{\it a} The "characteristic length scale" $r_0$ 
($\xi(r_0)\equiv1$) plotted as a function of the sample 
radius $R_s$ for various volume limited samples 
of  the Perseus-Pisces redshift survey. If the catalog is homogenous 
we should find a constant value for $r_0$. The fitting line has a
 slope of $D/6$ in agreement with a fractal behaviour. 
2{\it b} The "characteristic length scale" $r_0$ plotted as function 
of the absolute magnitude $M_{lim}$ 
 of the volume limited samples with different depth. To test the
 luminosity segregation hypothesis, one should find that $r_0$ is the 
same for sample with the same $M_{lim}$ and different depth. It is
 clear that there is no agreement between these values.

\item Fig. 3 The scaling exponents $\tau(q)$
as a function of the moment $q$ for the Perseus-Pisces redshift
survey (from Sylos Labini \etal 1996b). 
The multifractal behaviour is shown by the 
change of slope. For negative momenta the data are erratic because 
they are dominated by the smallest galaxies not present in the sample.

\item Fig. 4 {\it (a)} The first four iterations for the construction of
a multifractal binomial measure with $\mu_1=1/5$ and $\mu_2=4/5$.
{\it (b)} The same MF measure of {\it (a)} but after $15$ iterations.

\item Fig. 5 The $f(\alpha)$ spectrum for the binomial 
multifractal of Fig.4. The support of the 
measure distribution is compact in this case.

\item Fig. 6 {\it (a)} The measure distribution $P(\mu)$ 
(Eq.\ref{e38}) for the binomial multifractal of Fig.4.
{\it (b)} The tail of $P(\mu)$: the fitting curve is a power 
law with an exponential cut-off according 
to Eq.\ref{f9}. 
the exponent is $\delta =-1$ according to Eq.\ref{e40}.

\item Fig. 7
{\it (a)} A random multifractal in the one-dimensional Euclidean space
{\it (b)} A random multifractal in the two-dimensional Euclidean space

\item Fig. 8 The measure distribution $P(\mu)$
(Eq.\ref{e38}) for the binomial multifractal of Fig.7.
The fitting curve is a power
law with an exponential cut-off according
to Eq.\ref{f9}. the exponent is $\delta =-1.2$.

\item Fig.9 
 The scaling exponents $\tau(q)$
as a function of the moment $q$ for the Perseus-Pisces redshift
survey (from Sylos Labini \etal, 1996b).
In this case we have randomized the absolute magnitudes, and 
hence we have broken the correlation between galaxy luminosities
and positions.
 The multifractal behaviour is also broken and the 
reference line has slope $2$ in agreement with the simple
fractal properties.

\end{itemize}
\end{document}